\documentclass[12pt]{article}
\pdfoutput=1
\usepackage{draft}
\usepackage{multirow}
\usetikzlibrary{arrows.meta,bending}
\usepackage{colortbl}
\definecolor{darkgold}{RGB}{184,134,11}
\definecolor{bdry}{RGB}{204,224,250}
\newtheorem{theorem}{Theorem}

\newtheorem{claim}{Claim}

\begin{document}

\begin{titlepage}
\begin{center}

\hfill \\
\hfill \\
\vskip 1cm

\title{Construction of holomorphic SCFTs with minimal index}

\end{center}

\author{Du Pei}

\address{Centre for Quantum Mathematics, University of Southern Denmark, \\Campusvej 55, 5230 Odense, Denmark}

 \vfill

\begin{abstract} It has been an open question whether there exist holomorphic VOAs with $\mathcal{N}=1$ supersymmetry and Witten index 1 for any positive central charge $c$. The existence of such a theory with $c=288$ is closely related to the 576-periodicity of the TMF spectrum via the Segal--Stolz--Teichner conjecture. Here we report the construction of thousands of such theories by extending the approach initiated by Gaiotto and Johnson-Freyd via ternary codes.
  In addition, for every positive integer $k$, we also construct holomorphic $\mathcal{N}=1$ VOAs of central charge $c=12k$ with index $24/\gcd(24,k)$---conjectured to be the minimum---in a computer-aided search carried out by AI coding agents. We overcome the exponential-time barrier by using chains of smaller cells to build ternary codes, whose indices can then be evaluated via a transfer-matrix technique, and we use a symmetric orbifold construction to reduce the problem of $k=12$ and $24$ to that of $k=6$. Along the way, we also establish several results on ternary codes and their indices. Using an observation relating codes and Landau--Ginzburg orbifold models, we argue that the non-chiral doubles of the ternary code VOAs can always be gapped supersymmetrically. 

\end{abstract}
 \vfill

 \end{titlepage}

 \tableofcontents
\section{Introduction}

One surprising prediction of the Segal--Stolz--Teichner conjecture \cite{Segal,ST1,ST2} is that the elliptic genus of a 2d $(0,1)$ theory cannot be an arbitrary modular form, but has to live in the image of a ring homomorphism from topological modular forms,
\begin{equation}
    \pi_*(\TMF) \longrightarrow \mathrm{MF}_{*/2} .
\end{equation}
See \cite{GJ,GPPV,LP,Lin:Monster,AKL,PeiWu} for a small collection of works on tests of this constraint on physical theories and its applications, and \cite{hopkins2002algebraic,DFHH} for an overview as well as a comprehensive review of the subject of topological modular forms.

In particular, it predicts that a holomorphic SCFT $T$ of central charge $c=12k$ must have Witten index $\cI(T)$ satisfying
\begin{equation}
    k\cdot \cI(T) \equiv 0 \pmod{24}.
\end{equation}
Equivalently, the index $\cI$ must be divisible by $24/\gcd(24,k)$.

It has been an open question, first asked in \cite{GJ}, whether the minimum, 
\begin{equation}
\cI_{\rm min}=\frac{24}{\gcd(24,k)},
\end{equation} can always be attained by a holomorphic SCFT for arbitrary $k$.\footnote{\label{fn:conv}Our convention is that a holomorphic SCFT by definition is indecomposable with a unique vacuum in the Neveu--Schwarz sector. Therefore, it is basically synonymous with the term $\cN=1$ ``holomorphic (S)VOA,'' and we will not discuss extended supersymmetry in this work. Also, when regarding them as full-fledged physical theories, by default, we put their degrees of freedom in the right-moving sector so that the preserved supercharge is also right-moving (i.e.~naturally a 2d $(0,1)$ theory). This might be slightly unnerving for readers who prefer calling the right-moving sector ``anti-holomorphic'' and the left-moving sector ``holomorphic.'' Indeed, the $q$ powers of the elliptic genus for us would keep track of the momenta of left-moving/anti-holomorphic degrees of freedom, so we should perhaps use $\bar q$ instead. Luckily, for the theories we are going to consider, their elliptic genera have no holomorphic or anti-holomorphic $q$-dependence. Thus the confusion is minimal and a justifiable cost of using the author's idiosyncratic conventions.} One particularly interesting case is when $k\in24\bZ$, where the minimum positive value of $\cI$ is $1$. In this work, we answer this question in the affirmative by explicitly constructing such SCFTs.

\subsection{Status and strategy}
The case of $k=0$ is trivial, where one can take $T$ to be the trivial theory with a single bosonic massive vacuum. For $k=1$, the super-Moonshine module \cite{Duncan} has index 24, which is minimal. For $k=2$ through 5, the theories with minimal indices have been found by Gaiotto and Johnson-Freyd in the foundational work \cite{GJ}. Their approach is based on the construction of holomorphic SCFTs via self-dual ternary codes (reviewed in Section~\ref{sec:code}) and then gauging (a.k.a.~orbifolding) an internal discrete symmetry of the theory, as illustrated below
\begin{equation}
    \vcenter{\hbox{\fbox{\shortstack{self-dual ternary codes\\with automorphism}}}}\xrightarrow[\text{lattice VOA}]{\text{3-framed}}\vcenter{\hbox{\fbox{\shortstack{holomorphic SCFTs\\with symmetry}}}}\xrightarrow{\text{orbifolding}}\vcenter{\hbox{\fbox{\shortstack{holomorphic SCFTs\\with smaller index}}}}
\end{equation}

The approach taken in \cite{GJ} to find self-dual ternary codes with a desired automorphism is via a computer-aided random search that scales exponentially with $k$. Therefore, it would seem unrealistic to hope that one can eventually reach $k=24$ with $\cI_{\rm min}=1$---or even $k=12$ with $\cI_{\rm min}=2$---by brute force.  

Our solution to the $k=24$ problem is based on the observation that, in order to obtain such a theory with index 1, one just needs a theory with $k=6$ (i.e.~$c=72$) and index $\pm 4$, to which one applies a ``double symmetric square.''

\subsection{Symmetric squares}

Given a holomorphic SCFT $T$ with index $\cI$, one can construct a symmetric square of the theory
\begin{equation}
    \mathrm{Sym}^2(T):=(T\otimes T)/\mkern-7mu/\bZ_2
\end{equation}
where $\bZ_2$ acts by exchanging the two copies. The index of the new theory is 
\begin{equation}\label{ISym}
    \cI(\Sym(T))=\binom{\cI+1}{2}+\cI=\frac{\cI(\cI+3)}{2}.
\end{equation}
The two terms in the middle are indices of the untwisted and twisted sectors.\footnote{The simplest way to understand this formula is by considering the sigma model to $\cI$ points (counted with orientation), though this is not a holomorphic example. The untwisted sector is generated by maps from $S^1\sqcup S^1$---total space of the trivial $\bZ_2$ bundle over a circle---modulo the exchange symmetry (gauge transformation), and there are $\binom{\cI+1}{2}$ such maps (unordered pairs of possibly identical points). For the twisted sector, one considers maps from $S^1$---total space of the $\bZ_2$ bundle with non-trivial holonomy, which can be viewed as the boundary of a Möbius strip---and there are $\cI$ such maps. The equation \eqref{ISym} is also a special case of the Dijkgraaf--Moore--Verlinde--Verlinde formula for the elliptic genus of symmetric products \cite{DMVV}, adapted to the fermionic case and specialized to $S_N=\bZ_2$ and holomorphic theories which have no dependence on the modulus $\tau$ of $T^2$.}

This simple-looking formula has a surprising consequence: if one starts with a theory $T$ whose $\cI=-4$, then $T':=\Sym T$ will have index
\begin{equation}
    \cI(T')=\frac{-4\cdot(-4+3)}{2}=2,
\end{equation}
and its own symmetric square after stacking with the Arf theory, $T''=\Sym(T'\otimes\mathrm{Arf})$, has index
\begin{equation}
    \cI(T'')=\frac{-2\cdot(-2+3)}{2}=-1,
\end{equation}
which would then achieve our goal.

\begin{figure}[hbt]
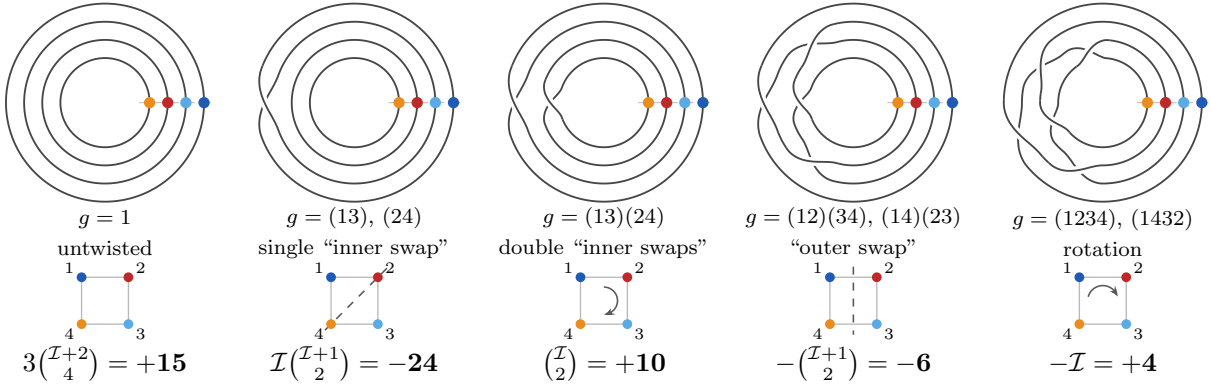
\centering

\caption{The sectors of the twisted $D_4$ orbifold of $T^{\otimes4}$ that gives the theory $T''$ with index $-1$, labeled by the conjugacy class of the holonomy $g\in D_4$ around the spatial circle. The four copies of $T$ (colored dots on the fiber over a point) are permuted by $g$, so going once around the circle joins them into the strands shown. The small square below each panel shows the conjugacy class of $D_4$ via its action on a square, with the four theories at the corners. Below each panel is the contribution of these sectors to the Witten index, as a polynomial in $\cI=\cI(T)$ and its value at $\cI=-4$. The total is thus $15-24+10-6+4=-1$. } \label{fig:D4}
\end{figure}

This can be summarized in the diagram below:
\begin{equation}\label{DoubleSym}
    \vcenter{\hbox{\fbox{\shortstack{{\large $T$}\\{\footnotesize (index $-4$)}}}}}\ \xrightarrow{\ \Sym\ }\ \vcenter{\hbox{\fbox{\shortstack{{\large $T'$}\\{\footnotesize (index $2$)}}}}}\ \xrightarrow{\ \otimes\,\mathrm{Arf}\ }\ \vcenter{\hbox{\fbox{\shortstack{{\large $-T'$}\\{\footnotesize (index $-2$)}}}}}\ \xrightarrow{\ \Sym\ }\ \vcenter{\hbox{\fbox{\shortstack{{\large $T''$}\\{\footnotesize (index $-1$)}}}}}
\end{equation}
Alternatively, an index-1 theory can also be achieved by a single twisted $D_4$ orbifold of a theory with index $\pm4$ as illustrated in Figure~\ref{fig:D4} and explained in Section~\ref{sec:symmetry}.

Here we have used the fact that, in the world of holomorphic SCFTs, there is a $\bZ_2$ involution\footnote{We will use $-1$ to denote the Arf theory and $-T$ for $T\otimes \mathrm{Arf}$, partially because of the relation to TMF and the convention there.}
\begin{equation}
    T\longrightarrow -T:=T\otimes \mathrm{Arf}
\end{equation}
given by stacking with an Arf theory---an invertible spin TQFT generating $\Omega^\Spin_2$---to flip the parity of Ramond-sector ground states, which then flips the sign of the Witten index. Therefore, getting a theory with index $-1$ is as good as a theory with index 1, and, for the ``seed'' theory $T$, having index 4 is as good as $-4$ as we can simply use $-T$ as the seed instead. In the diagram \eqref{DoubleSym}, we do need to use this operation to flip the middle theory $T'\to -T'$ as otherwise the index one gets would be 5 instead. Alternatively, one can combine the last two arrows as a twisted orbifold---discrete gauging but with a ``theta angle'' given by an invertible topological phase (or SPT).

Because of the existence of this double symmetric square, to find a holomorphic SCFT with minimal index for $k=12$ and $24$, it is good enough to find such a theory with $k=6$ and index $\cI=\pm 4$. As this is just slightly outside of the range studied in \cite{GJ}, it is actually possible via a brute-force random search with a required symmetry. To find more examples and, in particular, conceptually simple ones, we used agentic AI systems including GPT-6 Astra, GPT-6.1 Sol, Opus 5.5, and Fable 5.1. They produced and executed a collection of scripts in Python~3 using NumPy~\cite{Harris:2020xlr} to find thousands of such seed theories within a short time span of a week, as well as the minimal-index representatives for $k=8,9,10,11,13,14,15,16,18,21$. For $k=8$ and $16$, we use an orbifold construction by a new $Q_8$ symmetry.

The number of distinct codes found and the estimated average CPU time per code for each $k$ are documented in Table~\ref{tab:cpu}. 

\begin{table}[hbt]\centering
{\setlength{\tabcolsep}{3pt}
\begin{tabular}{c|c|c|c|c}
$k$ & codes & cell structure & candidates & seconds per code\\\hline
6 & $8{,}748$ & $(12,3)^3$ & $2\cdot10^{5}$ & $2.3$\\
7 & $82{,}188$ & $3\times(12,3)+(6,3)$ & $7\cdot10^{6}$ & $0.1$\\
8 & $4{,}884$ & $4\!\times\!(12,4)$ with $Q_8$ symmetry & $4\cdot10^{6}$ & $0.5$\\
9 & $407$ & $[(12,3)+(6,3)]^3$ & $2\cdot10^{7}$ & $40$\\
10 & $5{,}073$ & $5\times (12,3)$  & $6\cdot10^{6}$ & $1.5$\\
11 & $2{,}511$ & $5\times(12,3)+(6,3)$  & $8\cdot10^{6}$ & $5$\\
13 & $469$ & $6\times (12,3)+(6,3)$ & $9\cdot10^{6}$ & $40$\\
14 & $233$ & $7\times(12,3)$ & $7\cdot10^{6}$ & $70$\\
15 & $53$ & $[2\times(12,4)+(6,4)]^3$ & $2\cdot10^{7}$ & $\sim1000$\\
16 & $67$ & $8\!\times\! (12,4)$ with $Q_8$ symmetry & $6\cdot10^{8}$ & $\sim3000$\\
18 & $35$ & $[3\times(12,4)]^3$ & $2\cdot10^{8}$ & $\sim9000$\\
21 & $9$ & $[3\times(12,4)+(6,4)]^3$ & $10^{8}$ & $\sim50000$\\
\end{tabular}}
\caption{Yield of the computer search. The codes column counts the distinct codes found by the search. A ``candidate'' is a random choice of the cells in a chain (Figure~\ref{fig:ring}), and $(a,b)$ denotes a cell with $a$ total and $b$ boundary spins. ``$+$'' and ``$\times$'' join cells chosen independently, while an exponent repeats the block. For each $k$, a collection of different cell structures is often used in the search, while here we only list the one used for benchmarking. The candidates column counts those benchmarked, which the CPU time estimate is based on.}\label{tab:cpu}
\end{table}

\begin{figure}[hbt]\centering
\begin{tikzpicture}[scale=1.0, every node/.style={font=\small}]
  \useasboundingbox (-3.7,-3.4) rectangle (3.7,3.4);
  \def\L{6}\def\Rad{2.2}
  \foreach \j in {0,...,5} {
    \node[draw, rounded corners, fill=gray!12, minimum width=1.15cm, minimum height=0.8cm] (c\j) at ({90-360/\L*\j}:\Rad) {cell};
  }
  \foreach \j [evaluate=\j as \k using {int(mod(\j+1,6))}] in {0,...,5} {
    \draw[-{Stealth[length=2.4mm]}, thick] (c\j) to[bend right=18] node[midway, fill=white, inner sep=1pt] {$r$} (c\k);
  }
  \draw[-{Stealth[length=2.4mm]}, dashed, gray!70!black] ({90-360/\L*0.5+8}:{\Rad+1.05}) arc[start angle={90-360/\L*0.5+8}, end angle={90-360/\L*1.5-8}, radius={\Rad+1.05}] node[midway, anchor=south west, xshift=3pt, yshift=2pt, black] {shift symmetry};
  \node at (0,-0.1) {\begin{tabular}{c}length-$24L$ code \\with index\\ $\cI=\operatorname{Tr}(M^{L})$\end{tabular}};
\end{tikzpicture}
\caption{Structure of a ternary code presented as a closed chain of $L$  cells (with $L= 6$ in the figure). Each cell carries $12$ ternary ``spins'' and shares $r$ ``boundary spins'' with the next cell. The shift along the chain is the symmetry ($\bZ_6$ in the figure) to be gauged. The index of the code is then the trace of the $L$-th power of a $3^r\times3^r$ transfer matrix $M$, with the periodic (shown in the figure) or an anti-periodic boundary condition, as explained in Section~\ref{sec:index}.}\label{fig:ring}
\end{figure}
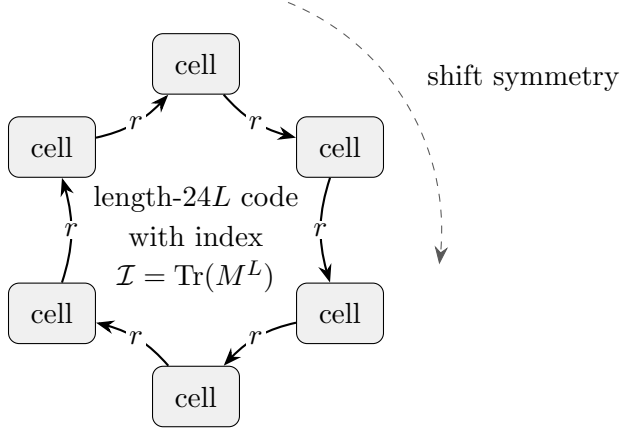

For higher values of $k$, the search was able to succeed by looking for ternary codes built from a closed chain of smaller ``cells'' (blocks of 6, 12 or 18 ``spins'' coupled only to their neighbors)---so that the index is a trace of a product of ``transfer matrices'' and can be evaluated rapidly (in milliseconds)---with the desired automorphism then being a shift along the chain. This is illustrated in Figure~\ref{fig:ring} with more details in later sections.

Since the index is multiplicative under tensor products, we can directly obtain minimal representatives saturating the TMF bound for $k=17=8+9$, $19=16+3$, $20=8+12$, $22=16+6$, $23=8+15$, and for every $k>24$ by tensoring with the index-1 theory at $k=24$. This proves
\begin{theorem}\label{thm:main} For every positive integer $k$,
there is a holomorphic $\cN=1$ VOA of central charge $c=12k$ with index $\frac{24}{\gcd(24,k)}$. 
\end{theorem}
The construction for each $k$ is summarized in Table~\ref{tab:constructions}.
\begin{table}[htb]\centering
\begin{tabular}{c|c|c|p{9cm}|c}
$k$ & $c$ & $\cI_{\min}$ & construction & gauged\\\hline
\rowcolor{blue!10} 6 & 72 & 4 & $(12,3)^3$, a Hadamard--Paley cell & $\bZ_6$ shifts\\
7 & 84 & 24 & $3\times(12,3)+(6,3)$ & --\\
8 & 96 & 3 & $4\times(12,4)$, $Q_8$-equivariant cells &  $Q_8$\\
9 & 108 & 8 & $(18,5)^3$, a Paley-type cell of width $18$ & $\bZ_3$ shifts\\
10 & 120 & 12 & $(12,4)^5$, a Hadamard--Paley cell & $\bZ_2=\langle-\mathds{1}\rangle$\\
11 & 132 & 24 & $5\times(12,3)+(6,3)$ & --\\
\rowcolor{teal!12} 12 & 144 & 2 & {$\Sym(T_6)$} &--\\
13 & 156 & 24 & $6\times(12,3)+(6,3)$ & --\\
14 & 168 & 12 & $7\times(12,3)$ & $\bZ_2=\langle-\mathds{1}\rangle$\\
15 & 180 & 8 & $[2\times(12,3)+(6,3)]^3$ & $\bZ_3$ shifts\\
16 & 192 & 3 & $8\times(12,4)$, $Q_8$-equivariant cells &  $Q_8$\\
17 & 204 & 24 & {$T_8\otimes T_9$}&--\\
18 & 216 & 4 & $[3\times(12,4)]^3$ & $\bZ_6$ shifts\\
19 & 228 & 24 & {$T_{16}\otimes T_3$}&--\\
20 & 240 & 6 & {$T_8\otimes T_{12}$}&--\\
21 & 252 & 8 & $[3\times(12,4)+(6,4)]^3$ & $\bZ_3$ shifts\\
22 & 264 & 12 & {$T_{16}\otimes T_6$}& --\\
23 & 276 & 24 & {$T_8\otimes T_{15}$}& --\\
\rowcolor{darkgold!22} 24 & 288 & 1 & {$\Sym(-\Sym(T_6))$, or a twisted $D_4$ orbifold of $T_6^{\otimes4}$}&--\\
$>24$ & --&-- & {$T_{k-24}\otimes T_{24}$}&--\\
\end{tabular}
\caption{Construction of holomorphic SCFTs with minimal index $\pm \cI_{\rm min}$ for each $k$ (such a theory is denoted as $T_k$). Cell structures listed can differ from the benchmarked ones in Table~\ref{tab:cpu} in several cases. Explicit generator matrices and verification records for every theory are available in \cite{ClaudeNotes}. The focus of this paper is the $k=6$ case (\textcolor{blue}{blue}) and the path to $k=12$ (\textcolor{teal}{teal}) and 24 (\textcolor{darkgold}{gold}) via its symmetric orbifolds.}\label{tab:constructions}
\end{table}

In the process, the AI agents found rich structures (e.g.~conceptually simple seed theories) and established various properties (e.g.~a proof that the $Q_8$ action used at $k=8,16$ is anomaly-free) to guide their search, and the interested reader is invited to check out the summary written by the agents themselves
\cite{ClaudeNotes}. It also links to a more agent-facing document with all the technical details.

 In this paper, we will focus on describing one of the $k=6$ theories below and the route to obtain an index-1 theory at $k=24$ from it. We start with a brief review of the construction of SCFTs via ternary codes in Section~\ref{sec:code}. We then describe the seed theory and a way to evaluate the index efficiently via a transfer-matrix technique in Section~\ref{sec:k6}. After that, in Section~\ref{sec:k24}, we discuss the discrete symmetries used in our approach and their gauging to obtain the $k=24$ theory. In the two remaining sections, we remark on how results for other values of $k$ are obtained (Section~\ref{sec:other}), and on the implications of our results for the geography of 2d supersymmetric theories, especially in relation to the Segal--Stolz--Teichner conjecture (Section~\ref{sec:SQFT}).

\section{Ternary codes and SCFTs}\label{sec:code}

A linear ternary code---in the sense of an error-correcting code---of length $n$ is a subspace $C\subset\mathbb{F}_3^{\,n}$, where $\mathbb{F}_3=\{0,+,-\}$ is the field of three elements. Elements in $C$ are called the ``codewords,'' and the dimension of $C$ as a linear space is referred to as the dimension of the code, $\dim( C)$. Therefore, there are in total $3^{\dim (C)}$ codewords.

One can choose $m$ linearly independent codewords to generate a code of dimension $m$. These are then the row vectors of a generator matrix $R\in\mathbb{F}_3^{\,m\times n}$, which, after row operations (i.e.\ a change of basis of $C$) and possibly a permutation of the coordinates, can be brought to the standard form $R=(\mathds{1}_m\,|\,A)$. The dual code is $C^\perp=\{y\in\mathbb{F}_3^{\,n}\ |\ x\cdot y=0,\ \forall x\in C\}$ with respect to the standard bilinear form, and $C$ is \emph{self-dual} if $C=C^\perp$. To have a self-dual code, $n$ has to be even with $m=n/2$, and, in the standard form, self-duality is the identity for the $m\times m$ matrix $A,$ 
\begin{equation}\label{eq:AAT}
A\,A^{T}=-\mathds{1}_{m}.
\end{equation}
Taking the determinant gives $\det(A)^2=(-1)^m$, and since $-1$ is not a square in $\mathbb{F}_3$, $m$ must be even, which requires that $n=2m$ is in fact a multiple of $4$ for a self-dual code of length $n$ to exist.

The smallest self-dual ternary code is the tetracode of length $4$, generated by $\left(\begin{smallmatrix}+&0&+&+\\0&+&+&-\end{smallmatrix}\right)$, while perhaps the best-known one is the ternary Golay code of length $12$, with generator matrix
{\setlength{\arraycolsep}{2.2pt}\small
\begin{equation}
R=\left(\begin{array}{cccccc|cccccc}+&0&0&0&0&0&0&+&+&+&+&+\\0&+&0&0&0&0&+&0&+&-&-&+\\0&0&+&0&0&0&+&+&0&+&-&-\\0&0&0&+&0&0&+&-&+&0&+&-\\0&0&0&0&+&0&+&-&-&+&0&+\\0&0&0&0&0&+&+&+&-&-&+&0\end{array}\right).
\end{equation}}%

One can turn a self-dual ternary code into an odd unimodular lattice \cite{ConwaySloane},
\begin{equation}
\Lambda(C)=\Bigl\{\tfrac{1}{\sqrt3}\,x:\ x\in\bZ^n,\ x \bmod 3\in C\Bigr\}\subset\mathbb{R}^n ,
\end{equation}
and the lattice (super) vertex operator algebra (VOA)  $V_{\Lambda(C)}$---related to the usual bosonic lattice VOA except that vectors of odd norm now create fermionic states---is a holomorphic VOA of central charge $c=n$. See \cite{Borcherds:1986,frenkel1988vertex,Dong:1993,Kac:1998} for standard facts about lattice VOAs.

Furthermore, this VOA has an $\cN=1$ supersymmetry and thus automatically defines an SCFT \cite{GJ}:  the $2n$ vectors $\pm\sqrt3\,e_i$ of norm $3$ give a ``3-frame'' $\sqrt3\,\bZ^n\subset\Lambda(C)$ from which one can construct a weight-$3/2$ field \begin{equation}
    G=\sum_i(\Gamma_{\sqrt3 e_i}+\Gamma_{-\sqrt3 e_i})
\end{equation} generating an $\cN=1$ superconformal algebra. Thus every self-dual ternary code of length $n=12k$ yields a holomorphic SCFT with $c=12k$, and a signed permutation of the coordinates (a monomial automorphism in coding-theory terminology, i.e.~a monomial matrix preserving $C$ \cite{HuffmanPless}) acts on the SCFT preserving $G$.

The Ramond ground states of $V_{\Lambda(C)}$ are labeled by the codewords of full support, $c=(c_1,\dots,c_n)$ with all $c_i\neq0$, and such a ground state is bosonic or fermionic according to the parity of the number of coordinates equal to $-$. The Witten index of the SCFT is then given by a combinatorial invariant of the code,
\begin{equation}\label{eq:index}
\cI\bigl(V_{\Lambda(C)}\bigr)=\mathrm{Index}(C):=\sum_{\substack{c\in C\\ \text{all } c_i\neq0}}\ (-1)^{\#\{i:\,c_i=-\}}.
\end{equation}
 It was conjectured for all $k$ and proved for even $k$ \cite{GJ} that $\mathrm{Index}(C)$ is divisible by $24$ for every self-dual code of length $12k$. We will also prove the case of odd $k$ and show that index $24$ is attained for every $k$ in Section~\ref{sec:other}.

\section{The $k=6$ theory with index 4}\label{sec:k6}

We can now start describing the theory with $k=6$ and index 4 from which we construct the $k=24$ theory with index 1.

\subsection{The seed code}\label{sec:seed}

The starting point (a ``seed'') of our construction is a code of length $72$ assembled from three copies of a $12$-spin ``cell.'' The cell is constructed with two $12\times12$ matrices over $\mathbb{F}_3$, with rows and columns labeled $0,1,\dots,11$,
{\setlength{\arraycolsep}{2.2pt}\small
\begin{equation}\label{eq:QS}
Q=\begin{pmatrix}
+ & + & + & \cellcolor{bdry}+ & 0 & \cellcolor{bdry}0 & 0 & 0 & \cellcolor{bdry}0 & 0 & 0 & 0 \\
+ & + & - & \cellcolor{bdry}- & 0 & \cellcolor{bdry}0 & 0 & 0 & \cellcolor{bdry}0 & 0 & 0 & 0 \\
+ & - & + & \cellcolor{bdry}- & 0 & \cellcolor{bdry}0 & 0 & 0 & \cellcolor{bdry}0 & 0 & 0 & 0 \\
+ & - & - & \cellcolor{bdry}+ & 0 & \cellcolor{bdry}0 & 0 & 0 & \cellcolor{bdry}0 & 0 & 0 & 0 \\
0 & 0 & 0 & \cellcolor{bdry}0 & + & \cellcolor{bdry}+ & + & + & \cellcolor{bdry}0 & 0 & 0 & 0 \\
0 & 0 & 0 & \cellcolor{bdry}0 & + & \cellcolor{bdry}+ & - & - & \cellcolor{bdry}0 & 0 & 0 & 0 \\
0 & 0 & 0 & \cellcolor{bdry}0 & + & \cellcolor{bdry}- & + & - & \cellcolor{bdry}0 & 0 & 0 & 0 \\
0 & 0 & 0 & \cellcolor{bdry}0 & + & \cellcolor{bdry}- & - & + & \cellcolor{bdry}0 & 0 & 0 & 0 \\
0 & 0 & 0 & \cellcolor{bdry}0 & 0 & \cellcolor{bdry}0 & 0 & 0 & \cellcolor{bdry}+ & + & + & + \\
0 & 0 & 0 & \cellcolor{bdry}0 & 0 & \cellcolor{bdry}0 & 0 & 0 & \cellcolor{bdry}+ & + & - & - \\
0 & 0 & 0 & \cellcolor{bdry}0 & 0 & \cellcolor{bdry}0 & 0 & 0 & \cellcolor{bdry}+ & - & + & - \\
0 & 0 & 0 & \cellcolor{bdry}0 & 0 & \cellcolor{bdry}0 & 0 & 0 & \cellcolor{bdry}+ & - & - & + \\
\end{pmatrix}\,,\qquad S=\begin{pmatrix}
0 & + & + & + & + & + & + & + & + & + & + & + \\
- & 0 & + & - & + & + & + & - & - & - & + & - \\
- & - & 0 & + & - & + & + & + & - & - & - & + \\
\cellcolor{bdry}- & \cellcolor{bdry}+ & \cellcolor{bdry}- & \cellcolor{bdry}0 & \cellcolor{bdry}+ & \cellcolor{bdry}- & \cellcolor{bdry}+ & \cellcolor{bdry}+ & \cellcolor{bdry}+ & \cellcolor{bdry}- & \cellcolor{bdry}- & \cellcolor{bdry}- \\
- & - & + & - & 0 & + & - & + & + & + & - & - \\
\cellcolor{bdry}- & \cellcolor{bdry}- & \cellcolor{bdry}- & \cellcolor{bdry}+ & \cellcolor{bdry}- & \cellcolor{bdry}0 & \cellcolor{bdry}+ & \cellcolor{bdry}- & \cellcolor{bdry}+ & \cellcolor{bdry}+ & \cellcolor{bdry}+ & \cellcolor{bdry}- \\
- & - & - & - & + & - & 0 & + & - & + & + & + \\
- & + & - & - & - & + & - & 0 & + & - & + & + \\
\cellcolor{bdry}- & \cellcolor{bdry}+ & \cellcolor{bdry}+ & \cellcolor{bdry}- & \cellcolor{bdry}- & \cellcolor{bdry}- & \cellcolor{bdry}+ & \cellcolor{bdry}- & \cellcolor{bdry}0 & \cellcolor{bdry}+ & \cellcolor{bdry}- & \cellcolor{bdry}+ \\
- & + & + & + & - & - & - & + & - & 0 & + & - \\
- & - & + & + & + & - & - & - & + & - & 0 & + \\
- & + & - & + & + & + & - & - & - & + & - & 0 \\
\end{pmatrix}\,.
\end{equation}}%
There are three ``boundary'' spins $\{3,5,8\}$, while the other nine are ``internal.'' The matrix $Q=H_4\oplus H_4\oplus H_4$ consists of three copies of the $4\times4$ Hadamard matrix on the diagonal blocks, and it is orthogonal over $\mathbb{F}_3$, 
\begin{equation}
QQ^T=4\cdot\mathds{1}=\mathds{1}.
\end{equation} The matrix $S$ is the skew Paley conference matrix.\footnote{With the coordinates $0,1,\dots,10,11$ read as the points $\infty,1,\dots,10,0$ of the projective line $\mathbb{P}^1(\mathbb{F}_{11})$, $S$ is the Paley matrix $S_{xy}=\chi(y-x)$, $S_{\infty y}=+$, $S_{x\infty}=-$, with $\chi$ the Legendre symbol. Its monomial automorphisms include $\mathrm{PSL}(2,11)$.} It is anti-orthogonal, 
\begin{equation}
SS^T=11\cdot\mathds{1}=-\mathds{1}.
\end{equation} 
There is one boundary spin in each Hadamard block. The columns of $Q$ and the rows of $S$ that belong to the boundary spins are shaded in blue.

From the cell we form the $12\times12$ matrices
\begin{equation}
D=Q\,P_{\rm in}\,S,\quad E=Q\,P_{\rm bd}\,S,\quad P_{\rm bd}=\mathrm{diag}(0,0,0,1,0,1,0,0,1,0,0,0),\quad P_{\rm in}=\mathds{1}-P_{\rm bd},
\end{equation}
so that $D+E=QS$, and the code $C$ is given by the generator matrix
\begin{equation}\label{eq:ring}
R=(\mathds{1}_{36}\,|\,A),\quad \text{with}\;\;
A=\begin{pmatrix} D & E & 0\\ 0 & D & E\\ -E & 0 & D\end{pmatrix}.
\end{equation}
The codewords are the row vectors $a\cdot R=(a\,|\,a\cdot A)$ with $a=(a_1,a_2,a_3)$, $a_1,a_2,a_3\in\mathbb{F}_3^{12}$, and
\begin{equation}
a\cdot A=\bigl(a_1D-a_3E,\ a_2D+a_1E,\ a_3D+a_2E\bigr).
\end{equation}
In other words, the block $a_{\alpha}$ contributes $a_{\alpha}D$ to the $\alpha$-th block and passes its three boundary spins $v_{\alpha}:=a_{\alpha}QP_{\rm bd}$ on to the $(\alpha+1)$-th one, through $a_{\alpha}E=v_{\alpha}\cdot S$, cyclically but with a sign when wrapping around, forming an ``anti-periodic spin chain.'' Although we will mostly use statistical mechanics terminology due to our background, such structures are studied extensively in coding theory, where such a system is known as a trellis \cite{BCJR}, here of ``tail-biting'' type \cite{CFV:tailbiting}.

What makes such a code desirable is that $A$ is ``almost block-diagonal,'' with the off-diagonal blocks given by the rank-$3$ matrix $E=QP_{\rm bd}S$. This makes the cells ``weakly coupled,'' which greatly facilitates the computations as we will explain later.

Before that, we still need to check self-duality \eqref{eq:AAT}. This is a straightforward computation using $QQ^T=\mathds{1}$, $SS^T=-\mathds{1}$ and $P_{\rm in}+P_{\rm bd}=\mathds{1}$: the diagonal blocks of $AA^T$ are 
\begin{equation}
    DD^T+EE^T=Q(P_{\rm in}+P_{\rm bd})Q^T\cdot(-1)=-\mathds{1},
\end{equation}
 while the off-diagonal ones vanish,
 \begin{equation}
 \pm DE^T=\mp\,QP_{\rm in}P_{\rm bd}Q^T=0.
 \end{equation}

The sign in the last row of $A$ does not affect self-duality---both the {periodic} chain (without the sign) and the {anti-periodic} chain are self-dual, and both are useful for different values of $k$. The $k=6$ seed just happens to be anti-periodic.

\subsection{Indices via transfer matrices}\label{sec:index}

The fact that the code \eqref{eq:ring} has index $-24$ can be checked directly by brute force, by going through the $3^{36}$ codewords (or the $2^{36}\approx7\times10^{10}$ words whose first 36 components are non-zero). However, as we are also interested in much larger values of $k$, a shortcut is needed, which is exactly what the block structure of \eqref{eq:ring} provides and what motivates the use of this very construction. 

Write a codeword of full support as $c=a\cdot R=(a_1,a_2,a_3\,|\,b_1,b_2,b_3)$ with $a_{\alpha}\in\{+,-\}^{12}$ and
\begin{equation}
b_{\alpha}=a_{\alpha-1}E+a_{\alpha}D=\bigl(v_{\alpha-1}+a_{\alpha}QP_{\rm in}\bigr)\,S,
\end{equation}
where, as defined before,
\begin{equation}
   v_{\alpha-1}:=a_{\alpha-1}QP_{\rm bd},
\end{equation}
supported on the three boundary coordinates, is the vector of boundary spins on the ``interface'' with block $\alpha-1$ (with $v_0=-v_3$ because of the anti-periodicity). The block $b_{\alpha}$ depends on $a_{\alpha}$ and the three spins $v_{\alpha-1}$ only, factorizing the index computation along the chain. Define the $27\times27$ ``transfer matrix,'' indexed by the $3^3$ possible values of the boundary spins,
\begin{equation}\label{eq:T}
M_{v,v'}=\sum_{\substack{a\in\{+,-\}^{12}\\ aQP_{\rm bd}=v'}} (-1)^{\#\{i:\,a_i=-\}}\;\sigma\bigl((v+aQP_{\rm in})S\bigr),
\end{equation}
where $a$ plays the role of $a_{\alpha}$, $v$ that of $v_{\alpha-1}$ and $v'$ that of $v_{\alpha}$, and
\begin{equation}
\sigma(b):=\begin{cases}(-1)^{\#\{i:\,b_i=-\}} & \text{if all } b_i\neq0,\\ 0 & \text{otherwise.}\end{cases}
\end{equation}
is the ``index'' of a length-$12$ block $b_{\alpha}$, so that the summand is the contribution of the length-$24$ word $(a_{\alpha},b_{\alpha})$ of one cell.\footnote{Notice that the labels $i$ and $\alpha$ are quite different, $\alpha=1,2,3$ is the label for the block, while $i=1,2,\ldots,12$ is the label for letters of words within a block.}

The transfer matrix $M_{v,v'}$ records, for an incoming state $v$ and an outgoing state $v'$, the signed number of local configurations that keep all $24$ coordinates of the block non-zero. Then
\begin{equation}\label{eq:trace}
\mathrm{Index}(C)=\operatorname{Tr}\bigl(JM^{3}\bigr),
\end{equation}
where $J_{v,v'}=\delta_{v',-v}$ implements the anti-periodic boundary condition. Each entry of $M$ is a sum over the $2^{12}$ local words $a\in\{+,-\}^{12}$ with full support, so the computation involves $27\cdot2^{12}\approx 10^5$ sign evaluations instead of $2^{36}$---a rather dramatic speedup.

In coding theory, this is the trellis evaluation of a weight enumerator \cite{Viterbi:1971,McEliece:BCJR}, in its tail-biting form \cite{Forney:duality}. Here we have adapted it to the signed count of full-support codewords.

Furthermore, once $M$ is known, the same cell repeated $L$ times gives a self-dual code of length $24L$ with index $\operatorname{Tr}(JM^{L})$ (or $\operatorname{Tr}(M^{L})$ for periodic chains) for every $L$. This is what overcomes the exponential-time barrier and makes the search for large $k$ possible.

For our seed theory, $M$ has $25$ non-zero eigenvalues and
\begin{equation}
\operatorname{Tr}(JM)=0,\qquad \operatorname{Tr}(JM^2)=-240,\qquad \operatorname{Tr}(JM^3)=-24 .
\end{equation}
The numbers on the right of these equations are deceptively small and hide dramatic cancellations between bosonic and fermionic states. In fact, dropping the signs in \eqref{eq:T} counts all Ramond ground states instead, and there are $83{,}000$ of them, with $41{,}488$ being bosonic and $41{,}512$ fermionic. 

The last equation means that the code is of index $-24$ and hence a legitimate candidate to be considered for discrete gauging. On the other hand, the first equality, $\operatorname{Tr}(JM)=0$, turns out to be a condition that makes it easier to control the index after orbifolding, and is not guaranteed to hold for arbitrary cells. In fact, the computer search screens candidates for exactly these two conditions: first the cheaper $\operatorname{Tr}(JM)=0$, then $\operatorname{Tr}(JM^3)=\pm24$ (or without the $J$ for periodic chains).

\subsection{Symmetry and orbifolding}\label{sec:symmetry}

The code \eqref{eq:ring} has an obvious symmetry---the shift along the anti-periodic chain
\begin{equation}
g:\ (a_1,a_2,a_3\,|\,b_1,b_2,b_3)\longmapsto(-a_3,a_1,a_2\,|\,-b_3,b_1,b_2),
\end{equation}
which is a signed permutation of the $72$ coordinates with $g^3=-\mathds{1}$, hence of order $6$. It acts on $V_{\Lambda(C)}$ preserving the $\cN=1$ structure, and we want to gauge it. 

In such a setting, i.e.~$C$ a self-dual code of length $12k$ with index $\pm24$ and $g$ a signed permutation of order $m=\gcd(k,24)$, one can gauge the $\bZ_m$ symmetry of $V_{\Lambda(C)}$ whenever the two conditions are satisfied \cite[Thm.~5.6]{GJ}: 
\begin{enumerate}
    \item[(i)] if $m$ is even, $g^{m/2}=-\mathds{1}$; 
    \item[(ii)] if $3\mid m$, then $g^{m/3}$ permutes the coordinates in $4k$ disjoint three-cycles and the codewords of full support fixed by $g^{m/3}$ are ``balanced,'' i.e.~with equal numbers of invariant bosonic and fermionic states. 
\end{enumerate}
Then the $\bZ_m$ action is non-anomalous and the orbifold $V_{\Lambda(C)}/\mkern-7mu/\bZ_m$ is a holomorphic $\cN=1$ SCFT with index $\pm24/m$.

Here $k=m=6$ and both hypotheses can be checked by inspection. For (i), $g^3=-\mathds{1}$ for our shift symmetry by construction. As for the first part of (ii), $g^2$ shifts by two blocks, $(a_1,a_2,a_3)\mapsto(-a_2,-a_3,a_1)$, so it permutes the $72$ coordinates in $24$ three-cycles. 

Codewords fixed by $g^2$ are those with $a_2=-a_1$, $a_3=a_1$ (and likewise for $b$)---the words that are constant along the chain up to an alternating sign. They are determined by a single cell closed on itself, which forms a length-$24$ self-dual code, with its signed count of full-support words being exactly the trace $\operatorname{Tr}(JM)=0$ seen previously. For our seed code above, the $80$ such words split evenly into $40$ bosonic and $40$ fermionic ones. 

The orbifold
\begin{equation}
T:=V_{\Lambda(C)}/\mkern-7mu/\bZ_6
\end{equation}
then has index
\begin{equation}
\cI(T)=\frac{-24}{6}=-4 ,
\end{equation}
saturating the TMF bound for $k=6$. It can therefore serve as a seed theory for our construction. Also notice that a theory obtained by gauging such a faithful anomaly-free discrete symmetry in our setting always has a unique NS vacuum---in this example, every twisted sector for the $\bZ_6$ symmetry in fact has ground-state energy at least $8/3$.

\section{The $k=24$ theory}\label{sec:k24}

We are now ready to construct the $k=24$ theory with index $\pm1$ from the seed theory above via symmetric orbifolding.

\subsection{The double symmetric square} 

In general, the permutation symmetry $S_N$ of $N$ copies of a theory can have an 't~Hooft anomaly which would then obstruct its gauging. However, in 2d, the anomaly for the $S_2=\bZ_2$ exchange symmetry only depends on the chiral central charge (or more generally on the gravitational anomaly of the 2d physical theory\footnote{In general, a 2d CFT with central charges $c_{\rm L}$ and $c_{\rm R}$ has gravitational anomaly $d=2(c_{\rm R}-c_{\rm L})$, normalized so that a single Majorana--Weyl fermion contributes $\pm1$. For an $\cN=(0,1)$ theory, $d$ is also the degree of its class in TMF via the Segal--Stolz--Teichner conjecture. A holomorphic $\cN=1$ SCFT of central charge $c$, viewed as the supersymmetric side of a $(0,1)$ theory, would then give a class in $\pi_{2c}(\TMF)$.})
and is given by $d\bmod 8=2c\bmod 8\in\bZ_8$ \cite{JF:MoonshineAnomaly,AKL},
which vanishes when $4\mid c$---all cases considered in this work satisfy this---making both symmetric squares in \eqref{DoubleSym} well defined. 

In general, there are two choices of a ``discrete theta angle'' (or a fermionic version of the more familiar ``discrete torsion'' \cite{Vafa:1986wx} in 2d) when gauging this $\bZ_2$, with the non-trivial one given by first stacking with the invertible $\bZ_2$-SPT phase $\mathrm{Arf}\otimes\mathrm{Arf}$. Notice that this would be a trivial phase if one forgets the $\bZ_2$ symmetry but a non-trivial one as a $\bZ_2$-symmetric phase (so genuinely an SPT phase). From the boundary point of view, the exchange acts on the two Majorana zero modes $\psi_1,\psi_2$ of $\mathrm{Arf}\otimes\mathrm{Arf}$ and maps the only possible mass term $i\psi_1\psi_2$ to $i\psi_2\psi_1$. However, as the two fields anti-commute, any $\bZ_2$-invariant mass term would have to vanish as $im\psi_1\psi_2=im\psi_2\psi_1=-im\psi_1\psi_2$.

When gauging the $\bZ_2$ exchange symmetry, this topological phase would flip the sign of the twisted-sector contribution and disallow symmetric states like $\ket{a}\otimes\ket{a}$ for bosonic $\ket{a}$ in the untwisted sector, and therefore the index becomes
\begin{equation}
\cI\longmapsto\frac{\cI(\cI+3)}{2}\qquad\text{and}\qquad \cI\longmapsto\frac{\cI(\cI-3)}{2}
\end{equation}
for the two choices respectively. The first is denoted by $\Sym(T)$ and the second by $\Sym(T\otimes\mathrm{Arf})=\Sym(-T)$, as we did in the introduction. Starting from our seed theory with index $-4$, to get the chain $-4\mapsto2\mapsto-1$, the second symmetric square is the twisted one. If one starts with an index-4 theory, both steps would need to be the twisted symmetric square.

\subsection{$D_4$ orbifold} 

One can also attempt to combine the two steps into a single one, by starting with four copies $T^{\otimes4}$ and gauging a subgroup of $S_4$, instead of first gauging the ``inner swaps'' $\bZ_2\times \bZ_2$ and then the ``outer'' $\bZ_2$. 

Notice that the two steps in the double symmetric square are now quite different. The inner $\bZ_2\times\bZ_2$ is an honest subgroup of $S_4$ generated by the transpositions $(13)$ and $(24)$, while the generator of the ``outer'' $\bZ_2$ has various lifts: it can be $(12)(34)$, $(14)(23)$, $(1234)$, or $(1432)$. These are the four lifts of the generator of the quotient in the extension,
\begin{equation}
1\longrightarrow\bZ_2\times\bZ_2\longrightarrow D_4\longrightarrow\bZ_2\longrightarrow1.
\end{equation}
The two lifts of order two, $(12)(34)$ and $(14)(23)$, split this extension, while the two rotations have order four. Together with the inner swaps, any of the lifts generates the same $D_4$ subgroup of $S_4$. Alternatively, it can be viewed as the wreath product $\bZ_2\wr\bZ_2=D_4\subset S_4$.

Gauging a normal subgroup and then the quotient is the same as gauging the extension. Therefore, to do it in one step, one should gauge the $D_4$ symmetry on $T^{\otimes4}$. This can be visualized as the symmetry group of a square whose corners carry the four copies of $T$, as illustrated in Figure~\ref{fig:D4}. The normal subgroup $\bZ_2\times\bZ_2$ is generated by the two diagonal reflections $(13)$ and $(24)$, whose product is the inversion (or $\pi$-rotation), while the outer swaps are the edge reflections $(12)(34)$ and $(14)(23)$, with the $\pm\frac\pi2$-rotations also in the same coset. 

The twist associated with the outer $\bZ_2$ now becomes a twist of the $D_4$ gauging via pullback along the quotient map, given by the character
\begin{equation}
    \chi:\quad D_4\longrightarrow D_4/(\bZ_2\times\bZ_2)\simeq\bZ_2,
\end{equation}
which is $-1$ exactly on the non-trivial coset. 

Recall that invertible fermionic phases with $G$ symmetry in 2d are classified by \cite{fidkowski2011topological,Kapustin:2014dxa}
\begin{equation}\label{fSPT}
\mathrm{fSPT}_{1+1}(G)
\simeq\Omega^\Spin_2(\rB G)^\vee
\simeq
\mathbb Z_2
\times H^1(G,\mathbb Z_2)
\times H^2(G,\U(1)),
\end{equation}
where the second identification holds as sets (under stacking, the three layers can form a non-trivial extension), and $\chi\in \Hom(D_4,\bZ_2)\simeq H^1(D_4,\mathbb Z_2)$ specifies the middle factor for $G=D_4$. 

The other generator for the middle factor $H^1(D_4,\bZ_2)\simeq \bZ_2^2$ can be taken to be the sign representation of $D_4\subset S_4$. Stacking a copy of Arf on each of the four $T$ factors gives exactly this character.

The first factor in \eqref{fSPT} is the familiar Arf theory---often excluded as an SPT since it does not require the $G$-symmetry. It can be used to flip the sign of the index of the resulting theory. We will need it at the end to go from $-1$ to $+1$. 

These three copies of $\bZ_2$ correspond to the choices in the double symmetric orbifold---one can stack an Arf theory at each step on $T$, $T'$ or $T''$. In other words, these are the three sign choices in
\begin{equation}
    T''=\pm\Sym(\pm\Sym(\pm T)).
\end{equation}

On the other hand, we do have the extra freedom to modify the discrete torsion $\eta\in H^2(D_4,\U(1))\simeq \bZ_2$, corresponding to some of the additional choices when gauging $D_4$ as a whole that the double symmetric orbifold cannot access.\footnote{\label{fn:dtorsion}In fact, it can be seen at the level of the normal subgroup $\bZ_2\times\bZ_2\subset D_4$ in the sense that its restriction to $\bZ_2\times\bZ_2$ is the familiar discrete torsion giving the phase $(-1)^{\langle a,b \rangle}$ on a torus with holonomies $a,b\in H^1(T^2,\bZ_2)$ for the two $\bZ_2$ factors paired via the intersection form. On a torus with $D_4$ holonomy $(g,h)$, it gives $-1$ exactly when $g\neq h$ are both non-trivial and generate a Klein four-group, and it is trivial on the rotation subgroup $\bZ_4$. }

Concretely, (a representative of) the class $\eta$ is a 2-cocycle
\begin{equation}
    \eta:\quad D_4\times D_4 \to \U(1),
\end{equation}
and on a torus with a flat $D_4$-connection, characterized by a pair of commuting holonomies $(g,h)$, it contributes a phase $\eta(g,h)/\eta(h,g)$ to the partition function of the SPT. 

Therefore, there are in total eight twists, obtained by first stacking $T^{\otimes4}$ with an invertible fermionic phase labeled by elements in $H^1(D_4,\bZ_2)\times H^2(D_4,\U(1))\simeq \bZ_2^3$ and then gauging.

In general, the index of a permutation orbifold $T^{\otimes n}/\mkern-7mu/G$ is
\begin{equation}\label{eq:permindex}
\cI\bigl(T^{\otimes n}/\mkern-7mu/G\bigr)=\frac{1}{|G|}\sum_{gh=hg}\varepsilon(g,h)\,\cI^{\,\#\text{orbits of }\langle g,h\rangle},
\end{equation}
a sum over commuting pairs, i.e.\ over $G$-bundles on $T^2$ with the non-bounding spin structure, in which the sign $\varepsilon(g,h)$ is the partition function of the invertible phase on that torus. In our case, we have
\begin{equation}\label{eq:epsilon}
\varepsilon(g,h)=(-1)^{q(\chi(g,h))}\,\nu_\chi(g,h)\,\frac{\eta(g,h)}{\eta(h,g)}.
\end{equation}
The first factor is given by the quadratic form $q$ on $H^1(T^2,\bZ_2)$ defined by the spin structure, evaluated at the image of $\chi$ on $(g,h)$. It can also be viewed as the difference between the Arf invariants of the untwisted spin structure and the one twisted by the $D_4$ holonomy via the $\bZ_2$-valued character $\chi$. The phase $\nu_\chi$ is a choice of convention for the origin of the space of discrete torsions.\footnote{When $\chi$ is non-trivial, it is better to regard $\eta$ as living in a torsor over $H^2(D_4,\U(1))$, as opposed to the latter space itself. Therefore, one has the freedom to choose the origin as part of the convention for each non-trivial $\chi$. Our choice is the following: we take $\nu_\chi=1$ if $\chi((13))=+1$, and $\nu_\chi$ the phase of the non-trivial discrete torsion (cf.~Footnote~\ref{fn:dtorsion}) if $\chi((13))=-1$. With this choice, $(-1)^{q(\chi)}\nu_\chi$ is the partition function of Arf stacked on each copy of $T$ if $\chi((13))=-1$ and on each copy of $T'$ if $\chi((12)(34))=-1$. This choice ensures that the four double symmetric orbifolds are exactly the twists with trivial $\eta$. }

Labeling a twist by $\bigl(\chi((13)),\chi((12)(34)),\eta\bigr)$, with $+$ for the trivial and $-$ for the non-trivial value, the eight twists give at $\cI=-4$
\begin{center}\small\setlength{\tabcolsep}{3.5pt}
\begin{tabular}{c|c|c|c|c|c|c|c|c}
twist & $(+,+,+)$ & $(+,+,-)$ & $(+,-,+)$ & $(+,-,-)$ & $(-,+,+)$ & $(-,+,-)$ & $(-,-,+)$ & $(-,-,-)$\\\hline
$\cI$ & $5$ & $-13$ & $-1$ & $-31$ & $119$ & $89$ & $77$ & $59$
\end{tabular}
\end{center}
(see also \cite{ClaudeNotes}). The four twists with trivial $\eta$, $(\pm,\pm,+)$, are the double symmetric orbifolds $\Sym(\pm\Sym(\pm T))$, with indices $5$, $-1$, $119$ and $77$. One of them indeed gives $|\cI(T'')|=1$, coming from the twist $(+,-,+)$, for which $\varepsilon(g,h)=-1$ exactly when the pair $(g,h)$ has a non-trivial image in $D_4/(\bZ_2\times\bZ_2)$. 

This is precisely the twisted double symmetric square, and its index is
\begin{equation}
\cI\bigl(T^{\otimes4}/\mkern-7mu/_{\!\varepsilon}D_4\bigr)=\frac{\cI^4+6\cI^3+3\cI^2-18\cI}{8}\,,
\end{equation}
which evaluates to $-1$ at $\cI=-4$. Grouping the commuting pairs by the conjugacy class of $g$ gives the five sectors of Figure~\ref{fig:D4}: the untwisted sector, the inner swaps, the double inner swap ($\pi$-rotation), the outer swaps, and the $\frac{\pi}{2}$-rotations respectively contribute
\begin{equation}
3\binom{\cI+2}{4},\qquad \cI\cdot\binom{\cI+1}{2},\qquad \binom{\cI}{2},\qquad -\binom{\cI+1}{2},\qquad -\cI.
\end{equation}
These can be obtained via a short computation, again pretending $T$ is the sigma model to $\cI$ points. One does need to note that in the sectors with $\chi(g)=+1$, the wavefunction should be antisymmetric under any $h\in D_4$ with $\chi(h)=-1$, while in the sectors with $\chi(g)=-1$ the states have reversed fermionic parity---the very reason for the minus sign in front of the last two cases---but are symmetric under any $h$. 

\subsection{States and fields}\label{sec:states}

The theories $T$, $T'$ and $T''$ involved in the double symmetric orbifold of a $k=6$ seed theory are explicit enough for their spectra to be computed exactly \cite{ClaudeNotes}. 

None of them has fields of weight $\tfrac12$ or $1$: the seed code has minimum distance $9$, so $\Lambda(C)$ has no vectors of norm $1$ or $2$, and the $72$ currents $\partial X_i$ are removed by orbifolding, since $g$ has no eigenvalue $1$. Each has $113$ fields of weight $\tfrac32$ (among them the supercurrent $G$), $2{,}928$ of weight $2$ and $42{,}001$ of weight $\tfrac52$. Below weight $3$ they are all in the untwisted lattice SVOA sector, while the first states from twisted sectors appear at weight $3$.

The NS character is a polynomial in the Hauptmodul 
\begin{equation}
t=(\vartheta_3/\eta)^{12}=q^{-1/2}+24+276\,q^{1/2}+2048\,q+\cdots
\end{equation}
of $\Gamma_\theta\subset \SL(2,\bZ)$, the index-$3$ subgroup generated by $\tau\mapsto\tau+2$ and $\tau\mapsto-1/\tau$ stabilizing the NS-NS spin structure. The polynomial has degree $c/12$, and for $T$,
\begin{equation}
Z^{\rm NS}_T=t^6-144\,t^5+6984\,t^4-129679\,t^3+751404\,t^2-629867\,t+13940,
\end{equation}
with its constant term being the number of ground states in the Ramond sector. These are far from unique, with about $5\times 10^{15}$ of them in the index-1 theory:
\begin{center}
\begin{tabular}{c|c|c|c|c}
theory & $c$ & $\cI$ & bosonic Ramond ground states & fermionic Ramond ground states\\\hline
$T$ & 72 & $-4$ & $6{,}968$ & $6{,}972$\\
$T'$ & 144 & $2$ & $48{,}587{,}870$ & $48{,}587{,}868$\\
$T''$ & 288 & $-1$ & $2{,}360{,}781{,}062{,}549{,}029$ & $2{,}360{,}781{,}062{,}549{,}030$
\end{tabular}
\end{center}
Such huge ground-state degeneracies, set against an index of $\pm1$, $\pm2$ or $\pm4$, are typical of the theories constructed in our approach. The small index comes from near-perfect cancellations between bosonic and fermionic ground states, which gives a sense of how deep in the ocean of holomorphic VOAs the theories with minimal index lie.

The same computation can be carried out for every seed code. For $k=6$, $Z^{\rm NS}_T$ is again a polynomial of degree $6$ in $t$ with leading terms $t^6-144\,t^5$, while the remaining coefficients can be used to distinguish different theories. Among the $8{,}748$ seed codes of Table~\ref{tab:cpu}, these polynomials take $4{,}579$ different values, giving rise to $4{,}579$ pairwise inequivalent theories $T$, and as many $T'$ and $T''$.\footnote{In fact, the first few terms in the $q$-expansion of $Z^{\rm NS}$ from $q^{-3}$ through $q^{-1/2}$ already distinguish $4{,}563$ out of the total $4{,}579$ characters.} The polynomial of every seed is listed in \cite{ClaudeNotes}. One can attempt to tell them apart further by finer invariants, such as the Lie algebra structure on the weight-$1$ fields. 

\section{Other values of $k$}\label{sec:other}

The search for minimal-index theories for other values of $k$ follows a similar strategy, with the specific constructions listed in Table~\ref{tab:constructions}. 

The yield decreases as $k$ becomes larger, mainly because the index-$24$ condition on the product of the cell transfer matrices becomes harder to satisfy as the distribution of the index becomes wider. As an example, at $k=21$ (chains of four cells with four boundary spins, repeated three times, $2\times10^4$ samples), the index of a random candidate has a median absolute value about $10^{8.6}$ and mean absolute value about $10^{11.6}$. Notice that these are still small when compared with the ground-state degeneracies above.

There are three main ingredients that, combined with a collection of observed structures, have greatly helped facilitate the search:

\paragraph{Tensor products.} This provides a shortcut to reduce the problem of finding minimal-index theories with $k=17,19,20,22,23$ and all $k>24$ to the rest of the values of $k$. It is rather convenient, as the search with these larger  $k$ values can be quite time-consuming. One might wonder whether there are other useful symmetric orbifolds, but unfortunately the two that we used to go from $k=6$ to $k=12$ and then to $k=24$ seem to be the only ones. The AI agents searched all relevant choices of subgroups of $S_N$ using GAP~\cite{GAP4}, with various twists, applied repeatedly together with tensor products, but found no other way to reach the minimal indices for $k\le24$.

\paragraph{Chains of cells with varying sizes.} For $k=7,9,10,11,13,14,15,18,21$, an index-$24$ code of length $12k$ was found as a closed chain of cells of width $6$, $12$ or $18$ in the sense of \eqref{eq:ring}, with the same or different cells in different positions but a common number of boundary spins, so that the index is a trace of a product of transfer matrices. When the TMF bound requires a symmetry of order $m=3$ or $6$ to saturate, the chain is built from a block of cells repeated three times and the balance condition is again a trace, while, for $m=2$, one simply gauges $g=-\mathds{1}$. At $k=18$ and $k=21$ the hits came from cells with four boundary spins ($81$-state transfer matrices), after the three-spin families had produced more than $10^8$ candidates without success. The search samples tuples of cells at random from pools of cells, with the traces computed in exact integer arithmetic (a quick screen modulo two primes, then the exact trace for the few survivors). Although hits are rare, at about one in $10^2$ candidates at $k=6,7$, one in $10^3$--$10^5$ for $k=9,\dots,14$, one in $10^6$ at $k=15$, and one in $10^7$--$10^8$ for $k=16,18,21$, each candidate costs only milliseconds to evaluate.

\paragraph{Quaternionic symmetry.} In our approach, for $k=8$ and $k=16$, saturating the bound by a code orbifold requires a group of order $8$, and no cyclic group was found: order-$8$ shifts are excluded by a $2$-adic theorem mentioned below, and every code with a signed-permutation symmetry $g$ of order $8$ with $g^4=-\mathds{1}$ that we examined has index divisible by $48$ rather than $24$. Instead, we use codes of length $96$ and $192$ with a quaternion group $Q_8$ symmetry. Since every pair of commuting elements of $Q_8$ generates a cyclic group and every non-identity element has $-\mathds{1}$ as a power, the argument of \cite[Thm.~5.6]{GJ} gives $\cI=24/8=3$ as soon as the $Q_8$ action is non-anomalous. For this construction to work, one does need to check that the action is anomaly-free, which is a non-trivial step, as part of the anomaly is invisible to genus-one data \cite{Yu:genusone}. The proof, available in \cite{ClaudeNotes}, connects the code at $k=8$ to the sum of $24$ tetracodes, through a chain of $23$ $Q_8$-equivariant neighboring codes. Each step is an orbifold by a $\bZ_3^2$ that preserves the $Q_8$ anomaly, and at the end one uses that the anomaly group $\mathrm{Hom}(\Omega_3^{\Spin}(\rB Q_8),\U(1))\simeq\bZ_8\oplus\bZ_4^2$ has exponent $8$ \cite{DGL}, which divides $24$---the number of free-fermion blocks. At $k=16$, a similar argument applies.

\paragraph{Other structures in ternary codes.} Several observations made along the way are documented in \cite{ClaudeNotes}, including:
\begin{itemize}
\item a $2$-adic theorem, $2^{v_2(L)+2}\mid\operatorname{Tr}(JM^L)$ for $L\ge4$, which rules out shift symmetries of order $8$ and $24$ in these chains, the reason why non-cyclic groups are used at $k=8,16,24$;
\item an explicit derivation of the index of the nine-state cells, and a ``selection rule'' (an empirical criterion to get a seed code) for the positions of the boundary spins.
\end{itemize}
Finally, two results settle Conjecture~1.4 of \cite{GJ}.
\begin{theorem}[divisibility]\label{thm:div24}
The index of every self-dual ternary code of positive length is divisible by $24$.
\end{theorem}
\noindent\emph{Sketch of proof.} Since the index of such a $C$ is divisible by 12 \cite[Thm.~1.3]{GJ} and is $0$ unless the length $n$ is a multiple of $12$, it suffices to show that it is also divisible by 8 when $12\mid n$. If two self-dual codes $C$ and $C'$ meet in a (non-self-dual) subcode $D$ of codimension one, a computation of the index of $D^\perp$ via Poisson summation gives (details in \cite[Sec.~2]{GJFConj14}) $$\mathrm{Index}(D^\perp)=3\cdot\mathrm{Index}(D).$$

On the other hand, the vectors in $D^\perp$ outside $C\cup C'$ have non-zero (squared) norm and cannot contribute to the index, as every vector of full support has norm $0 \pmod 3$. This gives
$$
\mathrm{Index}(D^\perp)=\mathrm{Index}(C)+\mathrm{Index}(C')-\mathrm{Index}(D),
$$
leading to
$$\mathrm{Index}(C)+\mathrm{Index}(C')=4\,\mathrm{Index}(D).$$  

The full-support words $\pm x$ of $D$ pair up with equal signs, so the index of $D$ is even and 
$$\mathrm{Index}(C')\equiv-\mathrm{Index}(C) \pmod 8.$$

Any two self-dual codes of the same length are joined by a chain of such neighbors \cite{CCMO:neighbors}, and a sum of tetracodes has index $0$, which proves the claim.\hfill$\square$

\begin{theorem}[existence]\label{thm:exist24}
For every $k\ge1$, there is a self-dual ternary code of length $12k$ with index $24$.
\end{theorem}
\noindent\emph{Sketch of proof.} For each $k$, one can build a chain out of $k$ cells each being a copy of the ternary Golay code ${\rm Go}_{12}$. With a suitable choice of gluing data, this code $C_k$ turns out to be self-dual with 
$$\mathrm{Index}(C_k)=(-1)^{k-1}\cdot\mathrm{Index}({\rm Go}_{12})=(-1)^k\cdot24,$$ and for odd $k$ negating one coordinate gives $+24$. The explicit construction of $C_k$ and a complete proof are in \cite[Sec.~3]{GJFConj14}.\hfill$\square$

Notice that these codes directly give the minimal-index theory when $\gcd(k,24) = 1$, and for $\gcd(k,24) = 2$ their orbifold by $\bZ_2=\langle-\mathds{1}\rangle$ has index $12$. But they do not cover the ``harder'' cases of $k=8$ and $16$ (for which we used a $Q_8$ symmetry) or $18$ and $21$ (the two largest values of $k$ below $24$ not reached by tensor products).

\section{The landscape of 2d theories}\label{sec:SQFT}

So far, we have found a collection of nice SCFTs, but they are still only a handful of landmarks in the vast landscape of 2d supersymmetric theories. What is their significance?

\subsection{Representing TMF classes}

A holomorphic SCFT with $k=24$ and index 1 is a ``periodic element'' in the sense that, under the Segal--Stolz--Teichner conjecture, it maps to $\Delta^{24}\in \pi_{576}(\TMF)$, which gives the 576-periodicity of TMF,
\begin{equation}
    \TMF[576] \xrightarrow{\,\cdot \Delta^{24}\,} \TMF.
\end{equation}
This is an equivalence with the inverse given by $\Delta^{-24}$, which can be represented by an anti-holomorphic theory obtained by simply putting the same theory in the left-moving sector without any right-moving degrees of freedom (cf.~the approach in \cite{LP}).\footnote{Previously, a theory mapping to $\Delta^{24}$ was known to exist in a slightly weaker sense \cite{AKL}. The $24$th tensor power $(V^{f\natural})^{\otimes24}$ of Duncan's super-Moonshine module (normalized to have index $-24$) has $c=288$ and Witten index $24^{24}$, while its $S_{24}$ permutation orbifold $(V^{f\natural})^{\otimes24}/\mkern-7mu/S_{24}$ has Witten index $-25{,}499{,}225$. As these two integers are coprime, there exist integers $p$ and $q$ such that $24^{24}\cdot p-25{,}499{,}225\cdot q=1$. Then the formal combination $p\,(V^{f\natural})^{\otimes24}\oplus q\,(V^{f\natural})^{\otimes24}/\mkern-7mu/S_{24}$ has index $1$ and lies in the periodicity class. However, such formal combinations are not holomorphic and do not have a unique NS ground state---the vacuum is massively degenerate---hence are not described by an SVOA. See also  \cite{TachikawaYamashita:Anderson,JohnsonFreydYamashita} for other related works on the interplay between quantum field theories and the 576-periodicity of TMF.}

Using this trick, we have

\begin{theorem}
    For every integer $k$, there is a holomorphically factorizable SCFT with $c_R-c_L=12k$ and index $\frac{24}{\gcd(24,k)}$.
\end{theorem} 
\noindent Here, stating it as ``holomorphically factorizable'' is a bit roundabout as one can always choose holomorphic representatives for $k>0$ and anti-holomorphic ones for $k<0$ by putting those in the left-moving sector, and, for $k = 0$, the trivial theory is in a sense both holomorphic and anti-holomorphic.

Another remark is that our construction via the double symmetric square is natural, as it is compatible with the $E_\infty$-ring structure of $\TMF$, where one also expects the very chain $-4\Delta^6\mapsto2\Delta^{12}\mapsto-\Delta^{24}$ via power operations. Details of this are spelled out in \cite[Sec.~I]{ClaudeNotes}, with assumptions on the existence and properties of certain equivariant Thom classes in equivariant elliptic cohomology.

Furthermore, as $\pi_*(\mathrm{MString}) \to \pi_*(\mathrm{tmf})$ is surjective \cite{Devalapurkar}, and $\TMF\simeq \mathrm{tmf}[\Delta^{-24}]$, we can find a 2d $(0,1)$ theory that conjecturally represents any class in $\pi_*(\TMF)$ by stacking the anti-holomorphic theories above with a 2d $(0,1)$ sigma model with a string manifold target. When the target is connected (which is always possible to arrange), it is reasonable to expect that, for a suitable choice of metric, the ground state in the NS sector is unique, and that, under the RG flow, the theory becomes an SCFT also with unique vacua (provided that the class is non-zero). Therefore, every non-zero TMF class is then expected to be represented by an indecomposable SCFT. 

\subsection{Paths to periodicity}

However, there are often multiple seemingly distinct realizations of the same TMF class, and it is a highly non-trivial question whether they are the same and how to actually deform one into another. 

In particular, the periodicity of TMF concerns the equation $\Delta^{24}\cdot\Delta^{-24}=1$, and a natural question is why this is true on the physics side---why a $c=288$ $\cN=1$ holomorphic SCFT $P$ with index 1 or its anti-holomorphic cousin $\bar P$ should be invertible. 

If one takes $\bar P\otimes P$, the non-chiral double of an index-1 holomorphic $\cN=1$ VOA we found, one obtains a 2d $\cN=(1,1)$ SCFT, and it seems far from obvious why this theory can be gapped in a way that preserves half of its supersymmetry. Indeed, such holomorphically factorizable SCFTs have no supersymmetric relevant operators, and a supersymmetric marginal operator would force the index to vanish.\footnote{A deformation preserving the right-moving supersymmetry has the form $\delta L=G_{-1/2}\,\mathcal{O}$ with $\mathcal{O}=A\otimes B$ of weights $(h+\tfrac12,h)$, with $B$ and $A$ coming from the holomorphic and anti-holomorphic factors. Its dimension is $2h+1$. Since every non-trivial field of a holomorphic VOA has weight at least $\tfrac12$, there cannot be such a relevant operator. On the other hand, a fermion field $\psi$ with $h=\tfrac12$ can lead to a marginal deformation, but its zero mode $\psi_0$ in the Ramond sector would cause the index to vanish. Therefore, in the theories we are considering with non-zero index, neither relevant nor marginal supersymmetric deformations can exist.}
Therefore, the theories that we want to gap are all SUSY-rigid and are already the endpoints of supersymmetric RG flows, and any supersymmetric deformations of them to the trivial theory must be large and non-perturbative. 

One approach is to first go ``upward'' to find a UV theory that can flow back to it and then, from that UV theory, deform to the trivial theory by gapping it (cf.~the approach in \cite{GJW} when studying theories representing the 0 class in TMF). This is in sharp contrast with the situation in 1d, where the 8-fold periodicity of KO-theory is associated with a quartic interaction for eight Majorana fermions, which is still a relevant deformation in one dimension \cite{Kitaev:2009mg,Fidkowski:2009dba}.

This observation seems to tell us that, to solve the periodicity problem, the natural first step is to find a suitable UV description for $\bar P\otimes P$. One very natural hope is to find a 2d $(1,1)$ sigma model that flows to $\bar P\otimes P$. Then, on the target, one can turn on a Morse function as a superpotential to gap the theory. 

Alternatively, there may be a better partner $P'$ for $\bar P$, such that $\bar P\otimes P'$ is easier to work with and can be the infrared limit of a 2d $(0,1)$ sigma model with a left-moving fermion bundle on the target. These theories can also be deformed by superpotentials, but it is harder to control phase transitions, which lead to wall-crossing in the parameter space (cf.~\cite{Gukov:2019lzi,Bao:2025xhl}).

However, exploring these directly for any $P$ with such a large central charge seems rather difficult. For example, one should expect that the target of the sigma model in the second scenario is at least 192-dimensional, since the central charge, equaling $\frac 32$ times the dimension in the UV, decreases along RG flows.

Our construction might provide a shortcut to achieve this by allowing one to work with simpler theories with $c=72$, via the factorization 
\begin{equation}
    \bar P\otimes P=(\bar V\otimes V)^{\otimes4}/\mkern-7mu/G, 
\end{equation}
with $V$ now a rather simple lattice SVOA but at the expense of having a very large $G=G_L\times G_R$ where each half is the wreath product $\bZ_6\wr D_4=\bZ_6^{\,4}\rtimes D_4$ of order $6^4\cdot8=10{,}368$ (with the twist described in Section~\ref{sec:k24} when orbifolding).

Therefore, the problem becomes whether one can gap $(\bar V\otimes V)^{\otimes4}$ in a $G$-equivariant way that preserves supersymmetry, a question closely related to what is known as symmetric mass generation (see \cite{Wang:2022ucy} for a recent review).

We will only make a baby step toward this goal, by arguing that, without worrying about preserving $G$, $(\bar V\otimes V)^{\otimes4}$ can be gapped. In fact, even a single copy can be gapped. In other words, we are showing ``$(\pm24\Delta^{6})\cdot(\pm24\Delta^{-6})=576$'' with the 24 as coefficients instead of exponents.

\subsection{UV theory of a ``single double''}

The lattice VOA $V_{\Lambda(C)}$ of a self-dual ternary code $C$ of length $n$ is a simple-current extension of $V_{\sqrt3\bZ}^{\otimes n}$. Each copy $V_{\sqrt3\bZ}$ is the chiral half of the ``$\U(1)_3$'' WZW model, which can be realized as the infrared limit of the $\cN=2$ Landau--Ginzburg model with superpotential $\Phi^3$.

The current $J=\tfrac{i}{\sqrt3}\,\partial X$ and the weight-$\tfrac32$ supercurrents $G^\pm=e^{\pm i\sqrt3X}$ generate the $\cN=2$ algebra with $c=1$. It is the $\ell=1$ member of the $\cN=2$ minimal series $c=\tfrac{3\ell}{\ell+2}$, whose $\ell$-th member is the infrared limit of a Landau--Ginzburg model with superpotential $W=\Phi^{\ell+2}$ \cite{Vafa:1988uu,Martinec:1988zu}.

There then must be a rather simple relation between $n$ copies of this minimal model and $\overline{V}_{\Lambda(C)} \otimes V_{\Lambda(C)}$. On the VOA side, the extension data is encoded in the ternary code $C$, while for the UV Landau--Ginzburg model, $C$ determines a subgroup of the $\bZ_3^n$ symmetry. Namely, each codeword $c$ acts on $\Phi_i$ by $\omega^{c_i}$ with $\omega=e^{2\pi i/3}$ a third root of unity. Therefore, the simplest and most natural lift of the doubled SCFT $\overline{V}_{\Lambda(C)} \otimes V_{\Lambda(C)}$ is the following. 

\begin{claim}\label{conj:LG}
For every self-dual ternary code $C$ of length $n$, the Landau--Ginzburg orbifold $\bigl(\sum_{i=1}^n\Phi_i^3\bigr)/\mkern-7mu/C$, with $C$ acting as a diagonal phase by $\Phi_i\mapsto\omega^{c_i}\cdot \Phi_i$, flows to $\overline{V}_{\Lambda(C)}\otimes V_{\Lambda(C)}$ in the infrared.
\end{claim}

\noindent The physical justification for this claim is the following. Using the well-tested relation between the UV LG model and the $\cN=2$ minimal model in the infrared, as well as the standard assumption that gauging a symmetry that is preserved along an RG flow leads to a pair of theories that are again connected by an RG flow, one can reduce the claim to a computation just in the IR SCFTs. On one side, we have a product of $n$ copies of the $\Phi^3$ minimal model, where the sectors are labeled by $\lambda\in\mathbb{F}_3^n$, with the same label on both sides. The orbifold by $C$ is expected to add twisted sectors in which the left and right labels can differ by an element of $C$, and project onto right labels in $C^\perp=C$. What remains matches exactly the sectors of $V_{\Lambda(C)}\otimes\overline{V}_{\Lambda(C)}$ on the other side. 

Claim~\ref{conj:LG} is reminiscent of Gepner models \cite{Gepner:1987vz,Gepner:1987qi} and of other Landau--Ginzburg orbifold constructions using minimal models \cite{Vafa:1989xc,Intriligator:1990ua}, which might suggest that it generalizes to codes that are not necessarily self-dual and even beyond ternary. 

First of all, a non-self-dual but self-orthogonal (i.e.~$C\subset C^\perp$) ternary code $C$ leads to a non-holomorphic VOA (with modules labeled by the cosets in $C^\perp/C$), but it can also be doubled via the diagonal pairing, and we claim that the doubled theory, $\overline{V}_{\Lambda(C)}\otimes_{C^\perp/C} V_{\Lambda(C)}$, is still the RG fixed point of a  $\bigl(\sum_{i=1}^n\Phi_i^3\bigr)/\mkern-7mu/C$ theory. A special case is the zero code $C = \{0\}$. Then $\Lambda(C) = (\sqrt3 \bZ)^n$, and $V_{\Lambda(C)} = V_{\sqrt3 \bZ}^{\otimes n}$. Its diagonal double is indeed the $n$ copies of $\ell=1$ minimal models. From this, one can add codewords to $C$ one by one, as long as they are linearly independent and orthogonal with themselves and each other, and recover the claim by ``gauging one $\bZ_3$ at a time.''\footnote{For the full theory, one can in fact go beyond self-orthogonal codes, as there is no obstruction to gauging the entire $\bZ_3^n$ symmetry. However, the orbifold theory obtained by gauging a general $C$ does not have a simple description in terms of a diagonal double of a code VOA. Viewing the double $\overline{V}_{\Lambda(C)}\otimes_{C^\perp/C} V_{\Lambda(C)}$ as a sandwich with an abelian Chern--Simons bulk and chiral and anti-chiral boundary conditions on the two sides, adding a new word to $C$ is the procedure of condensing the corresponding line operator in ${C^\perp/C}$. Self-orthogonality ensures that the line have trivial spin, while 
two words being orthogonal to each other ensures that two lines have trivial braiding phase---both are required for condensation. When $C$ is already self-dual, the bulk is invertible and the theory becomes holomorphically factorizable.} 

Then, it is also tempting to state a similar conjecture for $(\ell+2)$-nary codes (e.g.~codes over $\bZ_{\ell+2}$) with $\ell>1$, but there are two important differences. One is that the central charge of the minimal model is now $c=\tfrac{3\ell}{\ell+2}>1$, and the minimal model cannot be the doubled theory itself. Also, VOAs coming from $(\ell+2)$-nary codes generically don't have supersymmetry (see \cite{Kawabata:2023nlt,Kawabata:2023rlt} for related work).
Therefore, a natural conjecture seems to be that, for a self-orthogonal $(\ell+2)$-nary code $C$, the non-chiral double of its corresponding VOA can be obtained from a SUSY-breaking RG flow from the $\bigl(\sum_{i=1}^n\Phi_i^{\ell+2}\bigr)/\mkern-7mu/C$ theory. This also has a somewhat trivial generalization for $\ell=0$ (i.e. binary code), where the infrared limit of LG model---the ``$\ell=0$ minimal model''---is just a massive vacuum, and the conjecture should instead state that it can then ``flow up'' to the doubled theory in a non-supersymmetric way---almost a tautology. Therefore, the $\ell=1$ case (ternary code) is at the ``borderline'' when the central charge of the minimal model has increased from the $\ell=0$ case to match that of the VOA, and it will increase further for $\ell>1$ to live above the VOA.  

Finally, in Claim~\ref{conj:LG}, the orbifold is without discrete torsion, and it would be interesting to explore, if one allows it, what the corresponding theories are on the code-VOA side. 

\subsection{A gapping test}\label{sec:gapping}

The Landau--Ginzburg description in the UV makes it possible to gap the double via an $\cN=(1,1)$ superpotential---extra credit for preserving more supersymmetries than we need. 

The UV theory itself has the real superpotential $h_0=\mathrm{Re}\sum_i\Phi_i^3$, and we deform it by a $C$-invariant term,
\begin{equation}
\begin{aligned}
    h&=h_0+\epsilon\sum_i|\Phi_i|^2.
\end{aligned}
\end{equation}
For $\epsilon\neq0$, $h$ is a Morse function, resolving, for each $\Phi_i$, the degenerate critical point at $0$ into four non-degenerate ones at $$\{0,-\tfrac{2\epsilon}{3},-\tfrac{2\epsilon}{3}\omega,-\tfrac{2\epsilon}{3}\omega^2\}.$$

For $\epsilon>0$ the first is a minimum of $h$ (bosonic vacuum), fixed by $\bZ_3$, and the other three are index-1 saddle points (fermionic), permuted by $\bZ_3$. Turning on $\epsilon$ therefore deforms the theory into the $C$-orbifold of $n$ copies of a gapped theory with $1+3$ massive vacua. 

The elliptic genus is unchanged by this deformation and can be computed from the Witten index of the gapped theory. In our case, it is given by the average over pairs $(c,c')$ of the signed number of vacua fixed by both \cite{Dixon:1985jw,Dixon:1986jc,Vafa:1989xc,Hirzebruch:1990}. Here a field with $c_i=c'_i=0$ contributes $-2$, while any other field contributes $1$, from its fixed critical point $\Phi_i=0$ alone.\footnote{One quick way to see this is to notice that one can always set $c=0$ by modular transformations. Then, there is no holonomy on the spatial circle and there are $1+3$ states in the Hilbert space. If the holonomy $c'$ on the time circle is also zero, it contributes $1-3=-2$. But if the holonomy is non-trivial, only the $\bZ_3$ invariant bosonic state contribute, as the three fermionic ones are permuted and not contributing to the trace.
} This gives the elliptic genus of the LG orbifold model,
\begin{equation}\label{eq:LGindex} 
    \cI\left(\bigl(\sum_{i=1}^n\Phi_i^3\bigr)/\mkern-7mu/C\right)=\frac{1}{|C|}\sum_{c,c'\in C}(-2)^{\#\{i:\,c_i=c'_i=0\}}.
\end{equation}
This turns out to equal the index squared of $C$ for every self-dual code, 
\begin{equation}
    \mathrm{Index}(C)^2 =\frac{1}{|C|}\sum_{c,c'\in C}(-2)^{\#\{i:\,c_i=c'_i=0\}},
\end{equation}
 which is exactly the elliptic genus of $\overline{V}_{\Lambda(C)}\otimes V_{\Lambda(C)}$. Instead of proving the above identity here, we will actually work out the version refined by the left-moving R-symmetry $\U(1)_L$ and then also match the refined count in Appendix~\ref{app:refined}. 

Now that we know how to gap the $\overline{V}_{\Lambda(C)}\otimes V_{\Lambda(C)}$ theory, by first flowing upward to the Landau--Ginzburg orbifold model and then deforming the superpotential, one might wonder whether a similar technique works for $\bar T\otimes T$ or $\bar P \otimes P$. Unfortunately, this method does not directly extend to these cases, since the symmetries gauged in the construction of $T$, $T'$ and $P=T''$ act on one chirality only and are invisible in the Landau--Ginzburg description.

\paragraph{Acknowledgment.} We thank Theo Johnson-Freyd for discussion, and Ying-Hsuan Lin and Muyang Liu for collaborations on related research projects that in large part inspired the present approach. The work of the author is supported by research grant 42125 from VILLUM FONDEN and the Simons Collaboration on ``New Structures in Low-dimensional Topology.'' This material is based upon work supported by the National Science Foundation under Grant No.~DMS-2424139, while the author was in residence at the Simons Laufer Mathematical Sciences Institute (SLMath) during the Fall 2026 semester. We would also like to thank the staff at SLMath for great hospitality and for helping set up the computational resources on which our computer-aided search was run.

\appendix

\section{Refined elliptic genera}\label{app:refined}

The LG model has an $\cN=(2,2)$ supersymmetry, with a $\U(1)_L\times \U(1)_R$ R-symmetry. As $\U(1)_L$ commutes with the action of $C$ and fixes the right-moving supercharges, the elliptic genus of the orbifold can be refined by the left-moving R-charge $J_0$, whose fugacity we denote as $y:=e^{2\pi iz}$. 

For a single field with $W=\Phi^3$ (of R-charge $\tfrac13$), twisted by $\omega^a$ in space and by $\omega^b$ in time, the refined elliptic genus is  \cite{Witten:1993jg,Kawai:1993jk}
\begin{equation}\label{eq:Fab}
F_{a,b}(\tau,z)=y^{-a/3}\,\frac{\theta_1\bigl(\tau,-\tfrac{2z}{3}+\tfrac{a\tau+b}{3}\bigr)}{\theta_1\bigl(\tau,\tfrac{z}{3}+\tfrac{a\tau+b}{3}\bigr)}.
\end{equation}
At $z=0$ it reduces to the signed counts in \eqref{eq:LGindex}, with $F_{0,0}=-2$ and $F_{a,b}=1$ otherwise. The refined elliptic genus of the LG orbifold model is then
\begin{equation}
Z_{{\rm LG}/C}(\tau,z)=\frac{1}{|C|}\sum_{c,c'\in C}\ \prod_{i=1}^nF_{c_i,c'_i}(\tau,z).
\end{equation}
On the other side, the elliptic genus of $\overline{V}_{\Lambda(C)}\otimes V_{\Lambda(C)}$ refined by $J_0$, with $J=\tfrac{i}{\sqrt3}\sum_j\partial X_j$ acting on $\overline{V}_{\Lambda(C)}$, is the index of $C$ (from the right-moving side) times the Ramond-sector supertrace $\chi^{\rm R}(\tau,z)=\operatorname{Tr}_{\rm R}(-1)^Fq^{L_0-n/24}y^{J_0}$ of $\bar V_{\Lambda(C)}$ from the left-moving side. 

It turns out that the two agree for every self-dual ternary code $C$,
\begin{equation}\label{eq:refined}
Z_{{\rm LG}/C}(\tau,z)=\mathrm{Index}(C)\cdot\chi^{\rm R}(\tau,z).
\end{equation}
To see this, let $\psi_u$, $u\in\mathbb{F}_3$, be the same supertrace over the $V_{\sqrt3\bZ}$-module of label $u$, we have
\begin{equation}
\psi_u(\tau,z)=\frac{1}{\eta(\tau)}\sum_{\substack{m\in\bZ\\ m\equiv u\pmod 3}}(-1)^m\,y^{\frac{m+3/2}{3}}\,q^{\frac{(m+3/2)^2}{6}},
\end{equation}
with $u\in\{0,1,2\}$ standing for $\{0,+,-\}$. Let $\iota_u=\psi_u(\tau,0)$ be its Witten index with $\iota_0=0$ and $\iota_\pm=\pm1$. 

The quintuple product identity (see \cite{Cooper:2006quintuple} for a survey) gives, for each field, 
\begin{equation}
\frac13\sum_{b\in\bZ_3}\omega^{-b\kappa}F_{a,b}=-\psi_{2a+\kappa}\,\iota_{a+\kappa}.
\end{equation}
In $Z_{{\rm LG}/C}$, the sum over $c'$ projects onto $\kappa\in C^\perp=C$, and the change of variables $(c,\kappa)\mapsto(2c+\kappa,c+\kappa)$, which maps $C\times C$ onto itself, factorizes the sum into
\begin{equation}
\Bigl(\sum_{u\in C}\prod_i\psi_{u_i}\Bigr)\Bigl(\sum_{w\in C}\prod_i\iota_{w_i}\Bigr)=\chi^{\rm R}\cdot\mathrm{Index}(C).
\end{equation}
 At $y=1$, \eqref{eq:refined} gives the unrefined equality between the Witten indices found in Section~\ref{sec:gapping}. One can similarly show that---now working with the IR minimal model instead of the UV Landau--Ginzberg description---for the other three spin structures, the partition functions of the two theories are also equal. More details are given in \cite{ClaudeNotes}.

\providecommand{\doihref}[2]{\href{#1}{#2}}
\providecommand{\arxivfont}{\tt}

\bibliographystyle{JHEP} 
\bibliography{refs}

\end{document}